# On the stability of non-supersymmetric supergravity solutions

Ali Imaanpur[1] and Razieh Zameni[2]

Department of Physics, School of sciences
Tarbiat Modares University, P.O.Box 14155-4838, Tehran, Iran

**Abstract**

We examine the stability of some non-supersymmetric supergravity solutions that have been found recently. The first solution is $AdS_5 \times M_6$, for $M_6$ an stretched $CP^3$. We consider breathing and squashing mode deformations of the metric, and find that the solution is stable against small fluctuations of this kind. Next we consider type IIB solution of $AdS_2 \times M_8$, where the compact space is a $U(1)$ bundle over $N(1,1)$. We study its stability under the deformation of $M_8$ and the 5-form flux. In this case we also find that the solution is stable under small fluctuation modes of the corresponding deformations.

[1] aimaanpu@modares.ac.ir
[2] r.zameni@modares.ac.ir

## 1. Introduction

Selecting a stable solution among the many candidate supergravity solutions is a major problem in any Kaluza-Klein compactification. One way to guarantee the stability is to demand that the solution preserve a portion of supersymmetry [1-3]. In the absence of supersymmetry, on the other hand, it is difficult to conclude whether a particular solution is stable. In fact, one needs to examine the stability under small perturbations in all possible directions of the potential. Moreover, even if a solution is stable under such small perturbations, there still remains the question of stability under nonperturbative effects [4]. Finding non-supersymmetric stable solutions, however, becomes important if we are to construct realistic phenomenological models in which supersymmetry is spontaneously broken.

Freund-Rubin solutions can be divided into two main classes depending on whether or not the compact space encompasses (electric) fluxes [5, 6]. When the flux has components only along the $AdS$ direction, it has been observed that the majority of solutions either preserve supersymmetry (and hence stable), or at least are perturbatively stable. For solutions that support flux in the compact direction (Englert type), however, supersymmetry is often broken. They are in fact suspected to be unstable, though, the direct computation of mass spectrum and determination of stability is more involved. Englert type solution of $AdS_4 \times S^7$, for instance, was shown to be unstable [7], and this was further generalized to seven dimensional spaces which admit at least two Killing spinors [8]. Pope-Warner solution is another non-supersymmetric example which supports flux in the compact direction [9], and was proved to be unstable much later [10]. Englert type solutions, in spite of their possible instability, have played a key role in studying the holographic superconductors. By employing similar techniques that we use in this paper, domain wall solutions were found that interpolate between the Englert type and the skew-whiffed solutions. The domain wall solutions were then used to describe holographic superconductor phase diagrams [11].

The stability of Freund-Rubin type geometries of the form $AdS_p \times M_q$, where $AdS_p$ is anti-de Sitter spacetime and $M_q$ a compact manifold, has also increasingly been studied after the discovery of the AdS/CFT correspondence [12]. Stability is important for understanding a possible dual conformal field theory (CFT) description. For stable solutions, the spectrum of the masses directly yields the dimensions of certain operators in such a CFT. Unstable solutions can still have a dual CFT description but the physics is different [13]. Since the curvature of $AdS$ is negative, not all the tachyonic modes lead to instability. In fact, scalars with $m^2 < 0$ may also appear if their masses are not below a bound set by the curvature scale of $AdS$ [3].

Recently, some new non-supersymmetric compactifying solutions of eleven-dimensional supergravity and type IIB supergravity have been found [14, 15]. Specifically, the eleven-dimensional supergravity solution consists of $AdS_5 \times M_6$, where for $M_6$ there are two possible choices. For the first solution $M_6$ is $CP^3$ with the standard Fubini-Study metric, which was derived and studied in [16], and it was further shown that is perturbatively stable [17]. For the second solution $S^2$ fibers of $CP^3$ are slightly stretched with respect to the base manifold. Type IIB solution, on the other hand, is $AdS_2 \times M_8$, where $M_8$ is a $U(1)$ bundle over $N(1,1)$. All these solutions have fluxes in the compact direction, they break supersymmetry and therefore it is important to know whether they are stable.

It is also interesting to see how these new solutions might arise from near horizon geometries of some particular brane configurations. This would then lead us to the construction of the CFT duals [18]. For the eleven-dimensional supergravity solutions, first notice that the compact manifold admits a nontrivial 2-cycle over which we can wrap branes. Therefore, one way to get the $AdS_5$ factor is to construct a Ricci flat cone over the compact manifold and then consider fractional 3-branes (wrapped M5-branes over the 2-cycles) in the orthogonal directions placed at the tip of the cone. The near horizon geometry of this brane configuration would be $AdS_5 \times M_6$. Similarly, for the type IIB case, since $M_8$ is Einstein and admits nontrivial 3-cycles, we can construct a Ricci flat cone over it, and then put fractional D0-branes (D3-branes wrapped over 3-cycles of $M_8$) at the tip of the cone. Therefore we expect $AdS_2 \times M_8$ solution to arise as the near horizon limit of this D0-brane configuration.

In this paper we examine the stability of solutions under small perturbations of the metric. For getting consistent equations of motion on $AdS$, however, we also need to introduce deformations of the fluxes. Here we follow an approach which is close to that of [19, 20]. For compactification to $AdS_5$ the metric deformations correspond to the breathing and squashing modes. Including the deformation of the 4-form flux would correspond to three massive mode excitations on the $AdS$ space. In type IIB case, however, the bundle structure of the compact manifold allows a more general deformation, which, in turn, results in seven massive mode excitations. Apart from deriving the mass spectrum of small fluctuations, our approach has the advantage of providing us with a set of consistent reduced equations on $AdS$ space, so that any solution to these equations can be uplifted to a supergravity solution in eleven or ten dimensions.

## 2. Stability of $AdS_5 \times CP^3$ compactification

In this section we consider the solution $AdS_5 \times M_6$, where $M_6$ is $CP^3$ written as an $S^2$ bundle over $S^4$ [14], and study its stability under small perturbations. We start by deforming the metric along the fiber and the base by some unknown scalar functions on $AdS_5$. To get consistent reduced equations we see that the 4-form flux also needs to be deformed. After deriving the curvature tensor of the metric we write the supergravity equations of motion, and then linearize the equations around the known solutions. This allows us to read the mass of the small fluctuations corresponding to those deformations. If the mass squared falls in the Breitenlohner-Freedman range then the solution is stable against such perturbations.

To begin with, let us take the eleven dimensional spacetime to be the direct product of a 5 and 6-dimensional spaces,

$$ds_{11}^2 = ds_{AdS_5}^2 + ds_6^2. \tag{1}$$

For the 6-dimensional space the metric reads

$$ds_6^2 = d\mu^2 + \frac{1}{4}\sin^2\mu\, \Sigma_i^2 + \lambda^2(d\theta - \sin\phi\, A_1 + \cos\phi\, A_2)^2$$
$$+\lambda^2 \sin^2\theta (d\phi - \cot\theta(\cos\phi A_1 + \sin\phi A_2) + A_3)^2, \tag{2}$$

with $\lambda$ the squashing parameter, and

$$A_i = \cos^2\frac{\mu}{2}\Sigma_i, \tag{3}$$
$$d\Sigma_i = -\frac{1}{2}\epsilon_{ijk}\Sigma_j \wedge \Sigma_k. \tag{4}$$

This is an $S^2$ bundle over $S^4$, and for $\lambda^2 = 1$ we get the Fubini-Study metric on $CP^3$.

To discuss the stability, we deform the metric as follows:

$$d\bar{s}^2 = e^{2A(x)} g_{\alpha\beta} dx^\alpha dx^\beta$$
$$+ e^{2B(x)}(d\mu^2 + \tfrac{1}{4}\sin^2\mu\, \Sigma_j^2) + e^{2C(x)}(d\theta - \sin\phi\, A_1 + \cos\phi\, A_2)^2$$
$$+ e^{2C(x)}\sin^2\theta(d\phi - \cot\theta(\cos\phi A_1 + \sin\phi A_2) + A_3)^2, \tag{5}$$

where $g_{\alpha\beta}$ is the $AdS_5$ metric, and $A(x)$, $B(x)$, and $C(x)$ are arbitrary scalar functions on $AdS_5$. In fact, $B(x)$ and $C(x)$ correspond to what is usually called the breathing and the squashing mode deformations. We choose the following vielbein basis

$$\bar{e}^\alpha = e^{A(x)} e^\alpha \qquad \alpha = \bar{0}, \bar{1}, \bar{2}, \bar{3}, \bar{4}$$
$$\bar{e}^0 = e^{B(x)} e^0$$
$$\bar{e}^i = e^{B(x)} e^i \qquad i = 1,2,3$$
$$\bar{e}^a = e^{C(x)} e^a \qquad a = 5,6, \tag{6}$$

where the indices $\alpha, \beta, \ldots$ indicate the 5d spacetime coordinates, and the rest are related to the 6-dimensional space, and

$$e^0 = d\mu, \qquad e^i = \tfrac{1}{2}\sin\mu\, \Sigma_i,$$
$$e^5 = \lambda(d\theta - \sin\phi\, A_1 + \cos\phi\, A_2),$$
$$e^6 = \lambda\sin\theta\, (d\phi - \cot\theta\, (\cos\phi\, A_1 + \sin\phi\, A_2) + A_3). \tag{7}$$

Evaluation of the Ricci tensor of this deformed metric yields

$$\bar{R}_{\alpha\beta} = e^{-2A}\{R_{\alpha\beta} - \nabla^2 A \delta_{\alpha\beta} + 4\partial_\beta B \partial_\alpha(A-B) + 2\partial_\beta C \partial_\alpha(A-C)\}, \tag{8}$$
$$\bar{R}_{ij} = (3e^{-2B} - e^{2(C-2B)} - e^{-2A}\nabla^2 B)\delta_{ij}, \tag{9}$$
$$\bar{R}_{ab} = (e^{-2C} + e^{2(C-2B)} - e^{-2A}\nabla^2 C)\delta_{ab}. \tag{10}$$

Next, as in [14], we want to write a similar ansatz for the gauge field strength. However, since we have perturbed the metric with some scalar functions on $AdS$ space we must add an extra term for consistency. Further, it is easier first to write the Hodge dual ansatz as follows

$$\bar{*}_{11} F_4 = \bar{\epsilon}_5 \wedge (\alpha(x)\, e^{56} + \gamma(x)\, K) + \bar{*}_5\, d\eta \wedge Im\Omega, \tag{11}$$

where we have defined,

$$R_1 = \sin\phi(e^{01} + e^{23}) - \cos\phi(e^{02} + e^{31}), \tag{12}$$
$$R_2 = \cos\theta\cos\phi\,(e^{01} + e^{23}) + \cos\theta\sin\phi\,(e^{02} + e^{31}) - \sin\theta\,(e^{03} + e^{12}), \tag{13}$$
$$K = \sin\theta\cos\phi\,(e^{01} + e^{23}) + \sin\theta\sin\phi\,(e^{02} + e^{31}) + \cos\theta\,(e^{03} + e^{12}), \tag{14}$$

$$Re\Omega = R_1 \wedge e^5 + R_2 \wedge e^6, \tag{15}$$

$$Im\Omega = R_1 \wedge e^6 - R_2 \wedge e^5, \tag{16}$$

$$\omega_4 = e^0 \wedge e^1 \wedge e^2 \wedge e^3. \tag{17}$$

As $F_4 \wedge F_4 = 0$ (see (21)), the Maxwell equation reads

$$d \bar{*}_{11} F_4 = \bar{\epsilon}_5 \wedge (\alpha - \gamma) \wedge Im\Omega + d \bar{*}_5 d\eta \wedge Im\Omega = 0, \tag{18}$$

where we used [14],

$$de^{56} = Im\Omega, \quad dK = -Im\Omega, \quad dIm\Omega = 0. \tag{19}$$

Hence, Maxwell equation implies

$$\nabla^2 \eta = \gamma(x) - \alpha(x), \tag{20}$$

where $\nabla^2 = \bar{*}_5 d \bar{*}_5 d$. Changing the basis through (6), from (11) we see that

$$\bar{*}_{11} F_4 = \bar{\epsilon}_5 \wedge \left(\alpha e^{-2C(x)} \bar{e}^{56} + \gamma e^{-2B(x)} \overline{K}\right) + e^{-2B(x)-C(x)} \bar{*}_5 d\eta \wedge \overline{Im\Omega}, \tag{21}$$

where bar indicates barred basis in (6). Therefore, for $F_4$ we find

$$F_4 = -\alpha(x) e^{-2C(x)} \overline{\omega}_4 - \gamma(x) e^{-2B(x)} \overline{K} \wedge \bar{e}^{56} + e^{-2B(x)-C(x)} d\eta \wedge \overline{Re\Omega}. \tag{22}$$

Let us now check the Bianchi identity $dF_4 = 0$. Since $d\omega_4 = 0$, $dRe\Omega = 4\omega_4 - 2e^{56} \wedge K$, and also $Im\Omega \wedge K = Im\Omega \wedge e^{56} = 0$, the Bianchi identity requires

$$-d\left(\alpha e^{4B(x)-2C(x)}\right) = 4d\eta,$$

$$d\left(\gamma e^{2C(x)}\right) = 2d\eta. \tag{23}$$

The above equations, in turn, imply

$$\gamma(x) = -\frac{1}{2}\alpha(x) e^{4B(x)-4C(x)} + \frac{1}{2}\alpha_0 (e^{-2C_1} + 2e^{2C_1}) e^{-2C(x)}, \tag{24}$$

where $\alpha_0$ and $C_1$ are two constants. Using (20) and (24), the equation of motion for $\alpha$ reads

$$\nabla^2 \alpha = -4\alpha \nabla^2 B + 2\alpha \nabla^2 C - 4[\gamma - \alpha] e^{-4B+2C}, \tag{25}$$

as we will later expand around constant solutions, here we have dropped quadratic derivative terms.

Next, let us turn to the Einstein equations [14]:

$$R_{MN} = \frac{1}{12} F_{MPQR} F_N{}^{PQR} - \frac{1}{3.48} g_{MN} F_{PQRS} F^{PQRS}, \tag{26}$$

where $M, N, P, \ldots = 0, 1, \ldots, 10$. With ansatz (22), we can calculate the right hand side of the above equations:

$$\bar{R}_{ij} = (\tfrac{1}{3}\alpha^2 e^{-4C(x)} + \tfrac{1}{6}\gamma^2 e^{-4B(x)})\delta_{ij}, \tag{27}$$

$$\bar{R}_{ab} = (-\tfrac{1}{6}\alpha^2 e^{-4C(x)} + \tfrac{2}{3}\gamma^2 e^{-4B(x)})\delta_{ab}. \tag{28}$$

Using (9) and (10) on the LHS of the above equations yields

$$3e^{-2B} - e^{2(C-2B)} - e^{-2A}\nabla^2 B = \tfrac{1}{3}\alpha^2 e^{-4C} + \tfrac{1}{6}\gamma^2 e^{-4B}, \tag{29}$$

$$e^{-2C} + e^{2(C-2B)} - e^{-2A}\nabla^2 C = -\tfrac{1}{6}\alpha^2 e^{-4C} + \tfrac{2}{3}\gamma^2 e^{-4B}. \tag{30}$$

Combining (24), (25), (29) and (30), we get three equations which have a constant solution $\alpha_0^2 = 4$, $C_1 = B_1 = 0$ corresponding to the first solution in [14] with $\lambda^2 = 1$. To study the small fluctuations, we expand around this constant solution and only keep the linear terms to get

$$\begin{aligned}
\nabla^2 \alpha &= 22\alpha + \tfrac{176}{3}B - \tfrac{56}{3}C, \\
\nabla^2 B &= -\alpha + \tfrac{10}{3}B + \tfrac{14}{3}C, \\
\nabla^2 C &= 2\alpha + \tfrac{52}{3}B + \tfrac{8}{3}C.
\end{aligned} \tag{31}$$

Our next task is to find the mass spectrum. This is easily found by diagonalizing the mass matrix appearing on the RHS of equations (31). The mass spectrum reads

$$\mathbb{M} = diag(-2, 12, 18). \tag{32}$$

To see whether the first mode is stable, we need to invoke the Breitenlohner-Freedman (BF) stability bound on $AdS_{d+1}$ which requires

$$m^2 \geq m_{BF}^2 = -\tfrac{d^2}{4}, \tag{33}$$

for the mode to be stable. For $AdS_5$ we need to have $m^2 \geq -4$, so we conclude that the solution is stable against all three fluctuation modes. This conclusion agrees with the result of [17] who proved the stability of this particular solution by analyzing the spectrum of forms on $CP^3$.

The squashed solution in [14] with $\lambda^2 = 2$, on the other hand, here corresponds to a solution with $\alpha_0^2 = 4$, $B_1 = 0$, and $e^{2C_1} = 2$. Expanding and linearizing the three equations (25), (29) and (30) around this solution we find

$$\begin{aligned}
\nabla^2 \alpha &= 13\alpha - \tfrac{40}{3}B + \tfrac{232}{3}C, \\
\nabla^2 B &= -\tfrac{1}{4}\alpha + \tfrac{16}{3}B - \tfrac{1}{3}C, \\
\nabla^2 C &= \tfrac{1}{2}\alpha + \tfrac{16}{3}B + \tfrac{35}{3}C.
\end{aligned} \tag{34}$$

Diagonalizing the mass spectrum we find

$$\mathbb{M} = diag(3, 9, 18), \tag{35}$$

which is clearly stable.

## 3. Stability of type IIB compactifications to $AdS_2$

Another solution that we would like to study its stability is a compactification of type IIB theory on a $U(1)$ bundle over $N(1,1)$. Let us start by taking the following seven-dimensional metric of $N(1,1)$ [15]:

$$ds^2_{N(1,1)} = d\mu^2 + \tfrac{1}{4}\sin^2\mu(\Sigma_1^2 + \Sigma_2^2 + \cos^2\mu\,\Sigma_3^2) + \lambda^2(d\theta - \sin\phi\, A_1 + \cos\phi\, A_2)^2$$

$$+ \lambda^2 \sin^2\theta (d\phi - \cot\theta(\cos\phi\, A_1 + \sin\phi\, A_2) + A_3)^2 + \tilde{\lambda}^2 (d\tau - A)^2 , \tag{36}$$

where $\lambda$ and $\tilde{\lambda}$ are the squashing parameters, and

$$A_1 = \cos\mu\, \Sigma_1, \quad A_2 = \cos\mu\, \Sigma_2, \quad A_3 = \tfrac{1}{2}(1 + \cos^2\mu)\Sigma_3, \tag{37}$$

and,

$$A = \cos\theta\, d\phi + \sin\theta\, (\cos\phi\, A_1 + \sin\phi\, A_2) + \cos\theta\, A_3 . \tag{38}$$

Note that the base manifold admits a closed 2-form, the Kähler form:

$$J = \tfrac{1}{4} da = \tfrac{1}{4} d(\sin^2\mu\, \Sigma_3) = e^{03} - e^{12} , \tag{39}$$

so that $dJ = 0$. Therefore, we can construct a $U(1)$ bundle over $N(1,1)$ as follows

$$ds_8^2 = ds^2_{N(1,1)} + \hat{\lambda}^2 (dz - a)^2 , \tag{40}$$

with $\hat{\lambda}$ measuring the scale of new $U(1)$ fiber.

To discuss the small fluctuations, as in previous section, we perturb the metric by scalar functions on $AdS_2$ as follows:

$$d\bar{s}^2_{10} = e^{2A(x)} ds^2_{AdS_2} + e^{2B(x)}\left(d\mu^2 + \tfrac{1}{4}\sin^2\mu(\Sigma_1^2 + \Sigma_2^2 + \cos^2\mu\,\Sigma_3^2)\right)$$

$$+ e^{2C(x)} (d\theta - \sin\phi\, A_1 + \cos\phi\, A_2)^2$$

$$+ e^{2C(x)} (\sin^2\theta (d\phi - \cot\theta(\cos\phi\, A_1 + \sin\phi\, A_2) + A_3)^2)$$

$$+ e^{2E(x)} (d\tau - A)^2 + e^{2D(x)} (dz - a)^2 . \tag{41}$$

Let us choose the following basis,

$$\bar{e}^\alpha = e^{A(x)} e^\alpha \qquad \alpha = \tilde{0}, \tilde{1}$$

$$\bar{e}^i = e^{B(x)} e^i \qquad i = 0,1,2,3$$

$$\bar{e}^a = e^{C(x)} e^a \qquad a = 5,6$$

$$\bar{e}^7 = e^{E(x)} e^7, \quad \bar{e}^8 = e^{D(x)} e^8, \tag{42}$$

where,

$$e^0 = d\mu, \qquad e^1 = \frac{1}{2}\sin\mu\, \Sigma_1, \qquad e^2 = \frac{1}{2}\sin\mu\, \Sigma_2, \qquad e^3 = \frac{1}{2}\sin\mu\cos\mu\, \Sigma_3,$$

$$e^5 = (d\theta - \sin\phi\, A_1 + \cos\phi\, A_2),$$

$$e^6 = \sin\theta\, (d\phi - \cot\theta\, (\cos\phi\, A_1 + \sin\phi\, A_2) + A_3),$$

$$e^7 = d\tau - A, \quad e^8 = dz - a. \tag{43}$$

Now in terms of the barred basis the Ricci tensor is diagonal and reads,

$$\bar{R}_{ij} = \left(6e^{-2B} - 4e^{2(C-2B)} - 2e^{2(E-2B)} - 8e^{2(D-2B)} - e^{-2A}\nabla^2 B\right)\delta_{ij},$$

$$\bar{R}_{ab} = \left(4e^{2(C-2B)} - \frac{1}{2}e^{2(E-2C)} + e^{-2C} - e^{-2A}\nabla^2 C\right)\delta_{ab},$$

$$\bar{R}_{77} = 4e^{2(E-2B)} + \frac{1}{2}e^{2(E-2C)} - e^{-2A}\nabla^2 E,$$

$$\bar{R}_{88} = 16e^{2(D-2B)} - e^{-2A}\nabla^2 D. \tag{44}$$

To write the self-dual 5-form field strength ansatz, we follow the prescription presented in [15]. However, as the metric is deformed by scalar functions we need to add some extra terms. Let us start by writing the following 5-form

$$\omega_5 = (\alpha e^{-4B-D}\bar{\omega}_4 + \beta e^{-2B-2C-D}\overline{K} \wedge \bar{e}^{56} + \gamma e^{-2B-C-E-D}\bar{e}^7 \wedge \overline{Im\Omega}) \wedge \bar{e}^8$$

$$+ \xi e^{-2B-2C-E}\overline{J} \wedge \bar{e}^{567} + e^{-2B-C-D}d\eta_1 \wedge \overline{Re\Omega} \wedge \bar{e}^8$$

$$+ e^{-2C-E-D}d\eta_2 \wedge \bar{e}^{56} \wedge \bar{e}^{78} + e^{-2B-E-D}d\eta_3 \wedge \overline{K} \wedge \bar{e}^{78}, \tag{45}$$

with $\alpha$, $\beta$, $\gamma$, $\xi$, $\eta_1$, $\eta_2$, and $\eta_3$ are now taken to be scalar functions over spacetime and barred basis are defined as in (42). Requiring $\omega_5$ to be closed, we get

$$d\xi - 4d\eta_2 = 0,$$

$$-d\gamma - 2d\eta_2 + d\eta_3 = 0,$$

$$d\alpha - 8d\eta_1 - 4d\eta_3 = 0,$$

$$d\beta + 2d\eta_1 - 2d\eta_2 - d\eta_3 = 0, \tag{46}$$

where for $N(1,1)$ we used $de^{56} = 2\, Im\Omega$, $dK = -Im\Omega$, and $dRe\Omega = 8\omega_4 - 2e^{56} \wedge K$ (these are different from the ones in previous section as the bases are different). Solving the above equations for $\alpha$ we get

$$\alpha = -4\beta + 6\xi + 8\gamma + 4\beta_0 - 6\xi_0 - 8\gamma_0. \tag{47}$$

Now, taking the Hodge dual of (45) (with $\epsilon_{\tilde{0}\tilde{1}01235678} = 1$, where $\tilde{0}$ and $\tilde{1}$ refer to $AdS_2$ coordinates), we find

$$\bar{*}_{10}\,\omega_5 = (-\alpha e^{-4B-D}\bar{e}^{567} - \beta e^{-2B-2C-D}\bar{K}\wedge \bar{e}^7)\wedge \bar{\epsilon}_2$$

$$+(\gamma e^{-2B-C-E-D}\overline{Re\Omega} - \xi e^{-2B-2C-E}\bar{J}\wedge \bar{e}^8)\wedge \bar{\epsilon}_2$$

$$+e^{-2B-C-D}\,\bar{*}\,d\eta_1\wedge \overline{Im\Omega}\wedge \bar{e}^7 + e^{-2C-E-D}\,\bar{*}\,d\eta_2\wedge \bar{\omega}_4$$

$$+e^{-2B-E-D}\,\bar{*}\,d\eta_3\wedge \bar{K}\wedge \bar{e}^{56}. \tag{48}$$

Requiring the above 5-form to be closed results to the following equations

$$2\alpha e^{-4B+2C+E-D} + \beta e^{-2C-D+E} + 2\gamma e^{-D-E} - e^{2C-E-D}\nabla^2\eta_3 = 0,$$

$$4\beta e^{-2C-D+E} - 8\gamma e^{-D-E} + 8\xi e^{-2C-E+D} - e^{4B-2C-E-D}\nabla^2\eta_2 = 0,$$

$$2\alpha e^{-4B+2C+E-D} - \beta e^{-2C-D+E} - e^{E-D}\nabla^2\eta_1 = 0. \tag{49}$$

Using (46) and (49), we can find three equations

$$\nabla^2\xi = 16\beta e^{-4B+2E} - 32\gamma e^{-4B+2C} + 32\xi e^{-4B+2D},$$

$$\nabla^2\gamma = 2\alpha e^{-4B+2E} + \beta(e^{-4C+2E} - 8e^{-4B+2E})$$

$$+ \gamma(2e^{-2C} + 16e^{-4B+2C}) - 16\xi e^{-4B+2D},$$

$$\nabla^2\beta = 2\alpha(e^{-4B+2E} - 2e^{-4B+2C}) + \beta(e^{-4C+2E} + 2e^{-2C} + 8e^{-4B+2E})$$

$$+ 2\gamma(e^{-2C} - 8e^{-4B+2C}) + 16\xi e^{-4B+2D}. \tag{50}$$

With these constraints on scalar functions we can now write down the ansatz for the self-dual 5-form:

$$\bar{F}_5 = \bar{*}_{10}\,\omega_5 + \omega_5 \tag{51}$$

which satisfies the equation of motion $d*F_5 = 0$.

Let us now consider the Einstein equations, taking the dilaton and axion to be constant, in the Einstein frame we have [15]:

$$R_{MN} = \frac{1}{4\cdot 4!}\left(F_{MPQRS}F_N{}^{PQRS} - \frac{1}{10}F_{PQRSL}F^{PQRSL}g_{MN}\right) \tag{52}$$

$$+ \frac{e^{-\phi}}{4}\left(H_{MPQ}H_N{}^{PQ} - \frac{1}{12}H_{PQR}H^{PQR}g_{MN}\right) + \frac{e^{\phi}}{4}\left(F_{MPQ}F_N{}^{PQ} - \frac{1}{12}F_{PQR}F^{PQR}g_{MN}\right).$$

Using (44) and (51), the Einstein equations reduce to the following equations:

$$6e^{-2B} - 4e^{2(C-2B)} - 2e^{2(E-2B)} - 8e^{2(D-2B)} - e^{-2A}\nabla^2 B = \frac{\alpha^2}{4}e^{-8B-2D},$$

$$4e^{2(C-2B)} - \frac{1}{2}e^{2(E-2C)} + e^{-2C} - e^{-2A}\nabla^2 C = -\frac{\alpha^2}{4}e^{-8B-2D} + \frac{\beta^2}{2}e^{-4B-4C-2D} + \frac{\xi^2}{2}e^{-4B-4C-2E},$$

$$4e^{2(E-2B)} + \frac{1}{2}e^{2(E-2C)} - e^{-2A}\nabla^2 E =$$
$$-\frac{\alpha^2}{4}e^{-8B-2D} - \frac{\beta^2}{2}e^{-4B-4C-2D} + \frac{\xi^2}{2}e^{-4B-4C-2E} + \gamma^2 e^{-2E-4B-2C-2D},$$

$$16e^{2(D-2B)} - e^{-2A}\nabla^2 D =$$
$$\frac{\alpha^2}{4}e^{-8B-2D} + \frac{\beta^2}{2}e^{-4B-4C-2D} - \frac{\xi^2}{2}e^{-4B-4C-2E} + \gamma^2 e^{-2E-4B-2C-2D}. \tag{53}$$

The solution found in [15] corresponds to the following constant solution of equations (50) and (53):

$$\alpha_0 = \frac{3}{2}, \quad \beta_0 = \frac{3}{16}, \quad \gamma_0 = -\frac{3}{16}, \quad \xi_0 = -\frac{3}{8}, \quad e^{2C_1} = e^{2E_1} = \frac{1}{4}, \quad e^{2D_1} = \frac{3}{16}. \tag{54}$$

To discuss the stability of this solution, we linearize equations (50) and (53) about the above solution to get,

$$\nabla^2 B = 24B - 2C - E + 3D + 16\beta - 24\xi - 32\gamma,$$
$$\nabla^2 C = -4B + 26C + 5E - 3D - 32\beta + 48\xi + 32\gamma,$$
$$\nabla^2 E = -4B + 10C + 21E - 3D + 48\xi + 64\gamma,$$
$$\nabla^2 D = 12B - 6C - 3E + 21D - 48\xi,$$
$$\nabla^2 \beta = 3B - \frac{9}{2}C + \frac{15}{4}E - \frac{9}{4}D + 16\beta,$$
$$\nabla^2 \xi = 3C + \frac{3}{2}E - \frac{9}{2}D + 4\beta + 6\xi - 8\gamma,$$
$$\nabla^2 \gamma = -3B - \frac{3}{2}C + \frac{9}{4}E + \frac{9}{4}D + 16\gamma. \tag{55}$$

As in previous section, we can diagonalize the RHS of (55) to find the mass spectrum:

$$\mathbb{M} = diag(2.14, 2.14, 12, 24, 29.85, 29.85, 30), \tag{56}$$

with all the eigenvalues positive, we conclude that the solution (54) is stable against small fluctuations.

## 4. Conclusions

In this paper we examined the stability of some recently found non-supersymmetric solutions of ten and eleven dimensional supergravity. We perturbed the metric and the form fluxes by some space-time dependent scalar functions so that to reduce the equations of motion consistently on $AdS$ space. We then linearized these equations around solutions corresponding to those of [14] and [15]. For the compactification of the form $AdS_5 \times M_6$, we found that the two solutions, with squashing parameters of $\lambda^2 = 1$ and $\lambda^2 = 2$, are both stable against the kind of small fluctuations that correspond to the breathing and the squashing modes. This result is in agreement with the analysis of [17] who proved the stability in the case of $\lambda^2 = 1$ using a different approach. For type IIB solution of $AdS_2 \times M_8$, on the other hand, we observed that there are more modes that can be consistently excited on $AdS$ space. We derived the equations of motion of these modes, and by linearizing them around the solution of [15] showed that this solution is also stable.

We showed that the solutions of [14, 15] are stable against some particular small perturbations of the metric and the fluxes. However, to complete the proof of stability one needs to consider more general perturbations and study their spectrum. Moreover, having proved that a solution is perturbatively stable there still remains to check the solution against nonperturbative instabilities [4]. Recently, the authors of [21] showed that there is an instanton solution which destabilizes the $AdS_5 \times CP^3$ solution, and hence concluded that it is nonperturbatively unstable. It is therefore interesting to see whether similar instanton solutions exist for $AdS_2 \times M_8$.

The method we used in this paper led us to a set of consistent reduced equations on $AdS$ space. Consequently, a solution of these reduced equations can be uplifted to derive new eleven-dimensional or type IIB solutions. Therefore, apart from deriving the mass spectrum of small fluctuations, our approach can also be useful in searching for new supergravity solutions. In particular, it is interesting to look for domain wall solutions which interpolate between different vacua.